\documentstyle[aps,epsfig,amstex,twocolumn]{revtex}

\begin{document}
\title{Low energy monopole Modes of a Trapped atomic Fermi Gas}

\author{G.\ M.\ Bruun$^*$}
\address{Nordita, Blegdamsvej 17, 2100 Copenhagen, Denmark}
\maketitle

\begin{abstract} 
We consider the low energy collective monopole modes of a trapped weakly interacting atomic Fermi gas in the 
collisionless regime. The spectrum is calculated for varying coupling strength and chemical potential. 
Using an effective Hamiltonian,  we derive 
analytical results that agree well with numerical calculations in various regimes.
The onset of superfluidity is shown to lead to effects such as the vanishing of the energy required to create a 
Cooper molecule at a critical coupling strength and to the emergence of pair vibration excitations. Our analysis 
suggests ways to experimentally detect the presence of the superfluid phase in trapped atomic Fermi gases.
\end{abstract}

Pacs Numbers: 67., 05.30.Fk, 32.80.Pj 
\
\

The trapping and cooling of fermionic atoms is attracting increasing attention in the field
of ultra cold atomic gases. Temperatures as low as $T\sim0.2T_F$ with $T_F$ denoting the 
Fermi temperature have recently been observed experimentally for such gases~\cite{Experiments}. One of
the intriguing goals of the impressive experimental effort is to observe the theoretically 
predicted transition to a superfluid state below a critical temperature $T_c$ when the 
effective atom-atom interaction is attractive~\cite{StoofBCS}.

In this paper, the $T=0$ low energy collective monopole modes 
of a weakly interacting fermi gas in the collisionless limit are examined both 
analytically and numerically. We identify spectral 
features characteristic of the onset of superfluidity. Since the collective modes of trapped gases 
can be measured with high precision~\cite{Dalfovo}, this leads us to suggest appealing ways of experimentally 
observing the predicted phase-transition.

Consider a dilute gas of fermionic atoms of mass $m$ with an equal number of atoms $N_\sigma$ 
in two hyperfine states $|\sigma=\uparrow,\downarrow\rangle$
trapped in a spherically symmetric harmonic potential of the form $U_0(r)=m\omega_T^2r^2/2$. 
Assuming that the gas is in the dilute regime $k_F|a|\ll 1$ with $k_F=\sqrt{2m\mu_F}$ 
($\mu_F$ is the chemical potential), the interaction between atoms in the two different hyperfine states 
can be described as $g\delta({\mathbf{r}})$ where $g=4\pi a/m$ and $a$ is the $s$-wave scattering length.
We put $\hbar=1$ throughout the paper. 
For $a<0$ the effective atom-atom interaction is attractive and the gas is unstable toward the formation 
of Cooper pairs below a certain critical temperature $T_c$~\cite{StoofBCS}. 
The quasiparticle excitations of the superfluid state can be 
calculated by solving the Bogoliubov-de Gennes (BdG) equations~\cite{BruunBCS}. 

The collective excitations of the gas can be calculated in the collisionless limit
using the random-phase approximation (RPA). In a real space 
formalism, the poles of the matrix 
\begin{equation}\label{supermatrix}
\Pi(\omega) =\left\{
\begin{array}{cccc}
\langle\langle\hat{\rho}_\uparrow\hat{\rho}_\uparrow\rangle\rangle &
\langle\langle\hat{\rho}_\uparrow\hat{\rho}_\downarrow\rangle\rangle&
\langle\langle\hat{\rho}_\uparrow\hat{\chi}\rangle\rangle&
\langle\langle\hat{\rho}_\uparrow\hat{\chi}^{\dagger}\rangle\rangle\\
\langle\langle\hat{\rho}_\downarrow\hat{\rho}_\uparrow\rangle\rangle &
\langle\langle\hat{\rho}_\downarrow\hat{\rho}_\downarrow\rangle\rangle&
\langle\langle\hat{\rho}_\downarrow\hat{\chi}\rangle\rangle&
\langle\langle\hat{\rho}_\downarrow\hat{\chi}^{\dagger}\rangle\rangle\\
\langle\langle\hat{\chi}\hat{\rho}_\uparrow\rangle\rangle &
\langle\langle\hat{\chi}\hat{\rho}_\downarrow\rangle\rangle&
\langle\langle\hat{\chi}\hat{\chi}\rangle\rangle&
\langle\langle\hat{\chi}\hat{\chi}^{\dagger}\rangle\rangle\\
\langle\langle\hat{\chi}^\dagger\hat{\rho}_\uparrow\rangle\rangle &
\langle\langle\hat{\chi}^\dagger\hat{\rho}_\downarrow\rangle\rangle&
\langle\langle\hat{\chi}^\dagger\hat{\chi}\rangle\rangle&
\langle\langle\hat{\chi}^\dagger\hat{\chi}^{\dagger}\rangle\rangle
\end{array}
\right\}
\end{equation}
yield the modes of the system. Here,  $\langle\langle \hat{A}\hat{B}\rangle\rangle$ is the Fourier transform of the 
retarded function  $-i\Theta(t-t')\langle[\hat{A}({\mathbf{r}},t),\hat{B}({\mathbf{r}}',t')]\rangle$ 
and $\langle \ldots\rangle$ is the thermal average. Also, 
 $\hat{\rho}_\sigma({\mathbf{r}})=\hat{\psi}_{\sigma}^{\dagger}({\mathbf{r}})\hat{\psi}_{\sigma}({\mathbf{r}})$
and $\hat{\chi}({\mathbf{r}})=\hat{\psi}_\downarrow({\mathbf{r}})\hat{\psi}_\uparrow({\mathbf{r}})$. 
Correlation functions with the operator 
$\hat{\chi}({\mathbf{r}})$ are included in order to describe the effects of superfluidity.
A numerical scheme for calculating $\Pi(\omega)$  
within the RPA was described in Ref.\ \cite{BruunMottelson}.

The nature of the superfluid state depends on 
two parameters: the number of particles trapped and the interaction strength~\cite{BruunHeisel}. 
When the atom-atom interaction is sufficiently weak and/or there 
are not too many particles trapped, the gas is in the so-called intrashell regime; it can in many ways best 
be regarded as a giant nucleus. In this regime, the 
Cooper pairs are formed only between particles with angular quantum numbers $(l,m)$ and $(l,-m)$
residing in the same harmonic oscillator shell with radial quantum number $n$ [i.e.\ 
single particle energy  $\epsilon_n=(n+3/2)\omega_T$]. 
For stronger interactions and/or more particles 
trapped such that $\xi_{BCS}\ll R$ where $\xi_{BCS}=k_F(\pi m\Delta)^{-1}$ is the coherence length ($\Delta$ is the 
pairing gap) and $R$ is the size of the system given by the Thomas-fermi result $R\simeq\sqrt{2\mu_F/m}$,
 the superfluid state is best described as a quasi-homogenous bulk 
system. The transition between the two qualitatively different regimes is roughly determined 
by $\xi_{BCS}\simeq R$  or equivalently $\Delta\simeq\omega_T$.
 Using the bulk theory prediction $\Delta=(2/e)^{7/3}\mu_F\exp(-\pi/2k_F|a|)$~\cite{Gorkov} and 
the Thomas-Fermi result $\mu_F=(6N_\sigma)^{1/3}\omega_T$ (neglecting small effects due to the pairing 
as $\Delta/\mu_F\ll1$ in the dilute regime), the equation $\Delta=\omega_T$
yields $N_\sigma^*=\frac{1}{6}\left(\frac{e}{2}\right)^7\exp(3\pi/2k_F|a|)$. 
For a given coupling strength with $N_\sigma\ll N_\sigma^*$ atoms trapped, 
the gas is in the in the intrashell regime and for 
$N_\sigma\gg N_\sigma^*$ atoms trapped it is in the bulk regime.
With $k_F|a|=0.1$ this condition yields $N_\sigma^*=4\times10^{20}$ 
 and  for $k_F|a|=0.3$ we get $N_\sigma^*=9\times10^6$. Current experiments 
have $N\sim{\mathcal{O}}(10^5)$ particles trapped~\cite{Experiments}.
 Also, the present experimental effort uses 
optical traps as the confining potential as they allow greater greater flexibility 
with respect to which atomic hyperfine states can be trapped~\cite{Experiments}. The
optical traps used so far have $1.2$kHZ$\le\omega_T/2\pi\le12$kHZ
corresponding to temperatures $58$nK$\le T\le580$nK~\cite{Experiments}. Thus, the condition $k_BT_c\ll\omega_T$ 
is less restrictive than for magnetic traps. 
 Note also that $k_BT_c\simeq\Delta(T=0)/2$ in the intrashell regime 
can be orders of magnitude higher than the bulk 
prediction~\cite{BruunTc}.
The above estimates seem to indicate, that for the gas in the dilute limit it is not unlikely that future experiments on trapped 
superfluid fermionic atoms will be in the intrashell regime.

For $k_F|a|\gtrsim0.3$ the gas is not in the dilute limit. Recently, a number of papers have been dealing with the 
case of a Feshbach resonance mediated interaction, which is predicted to increase $T_c$ dramatically~\cite{Holland}.
This could make the experimental observation of the superfluid transition easier. It is however 
presently unclear what happens in this strongly correlated regime where issues such 
as the instability toward spinodal decomposition have to be addressed. 

The collective modes also depends qualitatively on which regime the gas is in. 
In the bulk regime, the modes can be calculated by looking at phase fluctuations of 
the pairing field. A gradient and frequency expansion of the lower right $2\times2$ part of Eq.\ (\ref{supermatrix})
then yields a hydrodynamic spectrum for the lowest modes of the gas~\cite{Baranov}. 

In light of the possible experimental relevance, we consider in this paper the collective modes in the intrashell regime. 
From the fact that there is only pairing between particles with 
quantum numbers $(n,l,m)$ and $(n,l,-m)$ follows a crucial simplification:
 Ignoring for the time being the effect of the Hartree field, the monopole pairing correlations
of the particles are accurately described by the effective Hamiltonian
\begin{gather}\label{EffectiveH}
\hat{H}_{eff}=\sum_{nlm\sigma}\xi_n \hat{a}^\dagger_{nlm\sigma}\hat{a}_{nlm\sigma}-\nonumber\\
\frac{2G\omega_T}{\Omega_{n_F}}\sum_{nlm,n'l'm'}(-1)^{m+m'}\hat{a}^\dagger_{nlm\uparrow}\hat{a}^\dagger_{nl-m\downarrow}
\hat{a}_{n'l'-m'\downarrow}\hat{a}_{n'l'm'\uparrow}
\end{gather}
in the sense that it can be shown that Eq.\ (\ref{EffectiveH}) reproduces the gap 
equation and thus the results concerning the quasiparticle properties presented in Ref.\ \cite{BruunHeisel}
in the intrashell regime. 
Here  $G=32\pi^{-2}k_F|a|/15$ is the effective coupling strength 
and $\Omega_{n_F}=(n_F+1)(n_F+2)/2$ is the pair degeneracy of the highest
harmonic oscillator level occupied with radial quantum number $n_F$.  The operator 
 $\hat{a}^\dagger_{nlm\sigma}$ creates a particle with quantum numbers $(n,l,m,\sigma)$ and energy 
$\xi_n=\epsilon_n-\mu_F$.
This effective Hamiltonian is much simpler to work with than the full Hamiltonian
and it allows several analytic results to be derived. It should be emphasized that Eq.\ (\ref{EffectiveH}) can be
derived from the full Hamiltonian in the intrashell regime;
 it thus describes the actual physical system accurately  and is not merely a schematic model as we shall see.

In order to derive analytical results, we write the RPA eqn.\ for $T=0$
in configuration space as~\cite{Blaizot}
\begin{equation}\label{RPAeq}
\left\{
\begin{array}{cc}
A&B\\
-B^*&-A^*
\end{array}
\right\}
\left\{
\begin{array}{cc}
X^\nu\\
Y^\nu
\end{array}
\right\}=
\omega_\nu\left\{
\begin{array}{cc}
X^\nu\\
Y^\nu
\end{array}
\right\}.
\end{equation}
Here $\omega_\nu$ is the frequency of the excitation 
 $|0\rangle\rightarrow |\nu\rangle=\hat{\Gamma}_\nu^\dagger|0\rangle$ 
where $|0\rangle$ and $|\nu\rangle$ is the RPA ground and excited state respectively. 
The matrices $A$ and $B$ are given by 
$A_{nlm,n'l'm'}=\langle BCS|\hat{\gamma}_{nl-m\downarrow}\hat{\gamma}_{nlm\uparrow}\hat{H} \hat{\gamma}^\dagger_{n'l'm'\uparrow}\hat{\gamma}^\dagger_{n'l'-m'\downarrow} |BCS\rangle$ 
and 
$B_{nlm,n'l'm'}=\langle BCS|\hat{\gamma}_{nl-m\downarrow}\hat{\gamma}_{nlm\uparrow}\hat{\gamma}_{n'l'-m'\downarrow}\hat{\gamma}_{n'l'm'\uparrow}\hat{H}| BCS\rangle$.
The superfluid ground state $|BCS\rangle$ and the quasi particle creation operator ${\hat{\gamma}}^\dagger_{nlm\sigma}$
with quantum numbers $(n,l,m,\sigma)$  can both can be obtained from a solution to the 
BdG equations~\cite{BruunBCS}.
Also, $X_{nlm}=\langle 0|\hat{\gamma}_{nl-m\downarrow}\hat{\gamma}_{nlm\uparrow}|\nu\rangle$ and 
$Y_{nlm}=\langle 0|\hat{\gamma}^\dagger_{nlm\uparrow}\hat{\gamma}^\dagger_{nl-m\downarrow}|\nu\rangle$. 
In the intrashell regime,  
$\hat{\gamma}_{nlm\sigma}=u_n\hat{a}_{nlm\sigma}\pm v_n\hat{a}^\dagger_{nl-m-\sigma}(-1)^m$, $u_n^2=(1+\xi_n/E_n)/2$
and  $v_n^2=(1-\xi_n/E_n)/2$ with $E_n=(\xi_n^2+\Delta^2)^{1/2}$~\cite{BruunHeisel}.
We have written the RPA equations appropriate for subspace spanned by monopole pairing excitations of the kind 
${\hat{\gamma}}^\dagger_{nlm\uparrow}{\hat{\gamma}}^\dagger_{nl-m\downarrow}+c.c.$ 

We first discuss the solutions to  Eq.\ (\ref{RPAeq}) when the gas is in the normal phase ($\Delta=0$).
Excitations of monopole symmetry are given by 
$\hat{\Gamma}^\dagger_\nu=\sum_{nlm}(-1)^m[X^\nu_n{\hat{\gamma}}^\dagger_{nlm\uparrow}{\hat{\gamma}}^\dagger_{nl-m\downarrow}+Y^\nu_n\hat{\gamma}_{nl-m\downarrow}\hat{\gamma}_{nlm\uparrow}]$,
 i.e.\ $X^\nu_{nlm}=(-1)^mX^\nu_n$ and $Y^\nu_{nlm}=(-1)^mY^\nu_n$ in Eq.\ (\ref{RPAeq}).
Using this we obtain after some algebra
\begin{equation}\label{NormalRPA}
\frac{1}{2G}=\sum_{n>n_F}\frac{1}{\omega+2\xi_n}-\sum_{n<n_F}\frac{1}{\omega+2\xi_n}.
\end{equation}
 Equation (\ref{NormalRPA}) determines the lowest collective monopole 
modes of the gas in the normal phase. These modes are not particle conserving. They correspond to adding/removing 
a pair of particles; i.e.\ one obtains for the addition mode
 $\hat{\Gamma}^\dagger_a=\sum_{n>n_Flm}(-1)^mX_n \hat{a}^\dagger_{nlm\uparrow}\hat{a}^\dagger_{nl-m\downarrow}
+\sum_{n<n_Flm}(-1)^mY_n{\hat{a}}^\dagger_{nlm\uparrow}{\hat{a}}^\dagger_{nl-m\downarrow}$ with energy $\omega_a$ 
and likewise for the removal mode.
The lowest particle conserving monopole excitation is then given by the state  
 $\hat{\Gamma}^\dagger_r\hat{\Gamma}^\dagger_a|0\rangle$ with energy $\omega_r+\omega_a$.

Consider the case $\mu_F=(n_F+2)\omega_T$ corresponding to a symmetric 
level distribution around the Fermi level with levels below $\mu_F$ (i.e.\ $n\le n_F$) completely filled.
In this case, the system is in the normal phase for $T=0$ for 
weak coupling strengths  such that $G<G_c$. $G_c$ is obtained by setting $\Delta=0$ in the gap equation 
reading  $G^{-1}-\sum_nE_n^{-1}=0$. This yields 
$G_c^{-1}\sim2^{5/2}\ln[4\exp(\gamma)n_F]/3$ where $\gamma\simeq0.577$ is Eulers constant~\cite{BruunHeisel}.
 Note that this equation is reproduced  by setting $\omega=0$ in Eq.\ (\ref{NormalRPA}).
We therefore conclude that the closed shell  system becomes unstable toward the formation of Cooper pairs when the 
energy cost of adding/removing a pair of particles goes to zero. 
Equation (\ref{NormalRPA}) can be solved for $G-G_c\rightarrow 0_-$ yielding
\begin{equation}\label{ARmodesGc}
\frac{\omega_r+\omega_a}{\omega_T}=2\left[\frac{2}{7\xi(3)}\right]^{1/2}
\sqrt{\frac{1}{G}-\frac{1}{G_c}},\ \ \ G-G_c\rightarrow 0_-
\end{equation}
with $\xi(3)=\sum_{n=1}^\infty n^{-3}\simeq1.202$. This equation  gives the frequency of the lowest 
particle conserving monopole mode as the closed shell system becomes unstable toward the formation of a 
superfluid with increasing coupling strength. For $G\rightarrow 0$, we can 
also solve Eq.\ (\ref{NormalRPA}) obtaining 
\begin{equation}\label{ARmodesG0}
(\omega_r+\omega_a)/\omega_T=2-4G,\ \ \ G\rightarrow 0.
\end{equation}
We have $\hat{\Gamma}^\dagger_a\hat{\Gamma}^\dagger_r\rightarrow \sum_{lm,lm'}(-1)^{m+m'}{\hat{a}}^\dagger_{n_F+1lm\uparrow}{\hat{a}}^\dagger_{n_F+1l-m\downarrow}\hat{a}_{n_Flm\uparrow}\hat{a}_{n_Fl-m\downarrow}$
for $G\rightarrow 0$; i.e.\ we simply take a pair of particles in the highest filled shell and put them into 
a monopole state in the lowest empty shell.  This obviously costs $2\omega_T$ in energy for $G\rightarrow 0$.
Equation (\ref{ARmodesG0}) can also be obtained by a  first
order perturbative calculation of the first excited state energy~\cite{Hogaasen}.
Equation (\ref{NormalRPA}-\ref{ARmodesG0}) give the lowest monopole excitation for a closed shell system 
in the normal phase.

We now solve Eq.\ (\ref{RPAeq}) when the gas is in the superfluid phase ($\Delta>0$). Using 
$X_{nlm}=(-1)^mX_n$ and  $Y_{nlm}=(-1)^mY_n$, we obtain after some rather lengthy algebra:
\begin{equation}\label{SuperRPA}
\left|
\begin{array}{cc}
G^{-1}+4\sum_n\frac{E_n}{\omega^2-4E_n^2} & 2\omega\sum_n\frac{\xi_n}{E_n(\omega^2-4E_n^2)}
\\
2\omega\sum_n\frac{\xi_n}{E_n(\omega^2-4E_n^2)} &G^{-1}+4\sum_n\frac{\xi_n^2}{E_n(\omega^2-4E_n^2)}
\end{array}
\right|=0.
\end{equation}
Equation (\ref{SuperRPA})
determines the frequencies $\omega$ of the lowest monopole pair vibrations of the gas in  the superfluid phase.
 First, we note that it follows from the gap equation that $\omega=0$ is a solution to Eq.\ (\ref{SuperRPA}).
This is the spurious solution coming from the fact that BCS theory breaks the $U(1)$ gauge symmetry 
of the pair-field or equivalently particle conservation. We now examine the two opposite cases of interest: The completely 
filled and the open shell case.

First consider the case of $\mu_F=(n_F+2)\omega_T$ with $G>G_c$ such that the gas is superfluid. The solutions to 
 Eq.\ (\ref{SuperRPA}) split into an even solution ($X_{n_F+n}=X_{n_F-n}$ and $Y_{n_F+n}=Y_{n_F-n}$) corresponding
 to the 
spurious mode with $\omega=0$ and an odd solution with $X_{n_F+n}=-X_{n_F-n}$ and  $Y_{n_F+n}=-Y_{n_F-n}$. Some algebra
yields the odd solution energy as
\begin{equation}\label{HogaasenGc}
\frac{\omega}{\omega_T}=\frac{2\Delta}{\omega_T}=\left[\frac{2^{1/2}3}{7\xi(3)}\right]^{1/2}
   \sqrt{\frac{1}{G_c}-\frac{1}{G}},\ \ \ G-G_c\rightarrow 0_+
\end{equation}
where we have used $\Delta=[2^{-3/2}7^{-1}\xi(3)^{-1}3]^{1/2}(G_c^{-1}-G^{-1})^{1/2}$ for $G>G_c$ in the 
closed shell case~\cite{BruunHeisel}.
 This is the pair vibration mode $\hat{\Gamma}_{P}^\dagger|0\rangle$ where the modulus $|\Delta|$ of $\Delta$
oscillates. Note that $2\Delta$ is smaller than  the lowest particle conserving quasiparticle excitation energy:
$2E_{n_F}=2\sqrt{\xi_{n_f}^2+\Delta^2}>2\Delta$ ($\xi_{n_F}=-\omega_T/2$). The pair vibration 
mode corresponds to coherent excitations of quasiparticle pairs of monopole symmetry such that the resulting 
energy is smaller than $2E_{n_F}$; the mode is strongly collective. It also conserves the number of particles 
and can  be excited by a particle conserving probe of monopole symmetry. We conclude that in 
the case of a closed shell  configuration, one can detect experimentally the onset of superfluidity with increasing coupling 
strength by  observing the lowest monopole excitation frequency of the gas vanish as given by Eq.\ (\ref{ARmodesGc}).
The system becomes increasingly unstable toward the excitation of pairs of particles into the lowest empty band 
as they 
can use the degeneracy of this band to maximize their spatial overlap and thereby increase the (absolute value of) 
interaction energy; the energy cost of forming a Cooper molecule simply 
goes to zero at the critical coupling strength. 
Once the system is superfluid for $G>G_c$, the 
lowest monopole excitation energy again increases as given by Eq.\ (\ref{HogaasenGc}). Now the ground state of 
the system already consist of correlated pair excitations across the Fermi level to create Cooper pairs 
with maximum spatial overlap and the lowest excitation is the  pair vibration.

In the case of an open shell with $\mu_F=(n_F+3/2)\omega_T$, the gas remains superfluid for 
$G\rightarrow 0_+$ for $T=0$. In this limit, there is only pairing 
in  the shell at the chemical potential ($n=n_F$)~\cite{BruunHeisel}. 
Solving the RPA equations for this shell, we obtain the $\omega=0$ spurious solution and the  pair breaking solution  
with energy $2E_{n_F}=2\Delta$ ($\xi_{n_F}=0$). Residual interactions between 
between the excitations lower the frequency slightly below $2\Delta$. 
To a good approximation however, the lowest monopole  excitation energy of the system is
\begin{equation}\label{OpenRPA}
\frac{\omega}{\omega_T}\simeq\frac{2\Delta}{\omega_T}=\frac{2G}{1-2^{5/2}\ln(e^\gamma n_F)G/3},\ \ \ G\rightarrow 0_+
\end{equation}
where we have used 
$\Delta/\omega_T=G[1-2^{5/2}\ln(\exp(\gamma) n_F)G/3]^{-1}$ in the open 
shell case~\cite{BruunHeisel}. Superfluidity can thus also in the open shell case be detected by the presence of a low 
energy $\omega\simeq2\Delta$ mode.

\begin{figure}
\centering
\epsfig{file=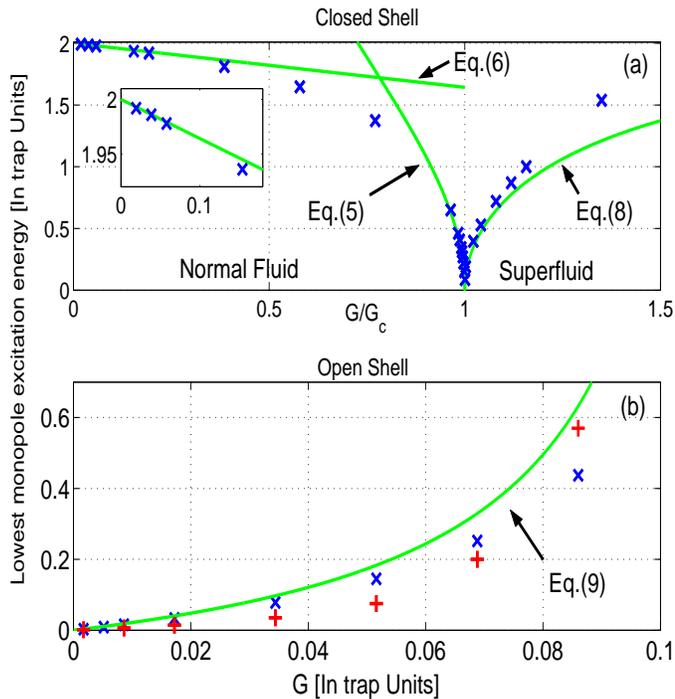,height=0.4\textheight,width=0.5\textwidth,angle=0}
\caption{The lowest monopole excitation  energy in units of $\omega_T$ for  (a): a closed shell system as a function of 
$G/G_c$ and (b): an open shell system as a function of $G/\omega_T$. The inset in (a) shows the $G\rightarrow0$ region in detail.}
\label{PairVibfig}
\end{figure}

We now compare our analytical results with an exact numerical solution of the RPA equations obtained using the method 
described in Ref.\ \cite{BruunMottelson}. First, consider the case of a closed shell system 
 [$\mu_F=(n_F+2)\omega_T$]. In fig.\ \ref{PairVibfig} (a)
we plot the lowest monopole excitation energy as a function of the coupling strength $G/G_c$. For $G<G_c$ the 
excitation is $\hat{\Gamma}^\dagger_r\hat{\Gamma}^\dagger_a|0\rangle$ whereas for $G>G_c$ the excitation is 
 $\hat{\Gamma}_{P}^\dagger|0\rangle$. 
The $\times$'s are obtained from a numerical calculation of the 
matrix given in Eq.~(\ref{supermatrix}) and the lines are Eq.~(\ref{ARmodesGc}), (\ref{ARmodesG0}), and (\ref{HogaasenGc}).
We have chosen $n_F=50$ corresponding to
$\sim2\times10^4$ particles trapped in each of the two hyperfine states. There is very good agreement between the
numerical results and the analytical predictions in their regions of validity. This shows that effective Hamiltonian given 
by Eq.~(\ref{EffectiveH}) accurately describes the pairing correlations for the physical system 
in the intrashell regime. As the coupling strength 
increases, the intrashell pairing ansatz eventually breaks down; this occurs  for $G/G_c\sim 1.2$ 
for this set of parameters~\cite{BruunHeisel}.
 For stronger coupling, 
Eq.~(\ref{EffectiveH}) is invalid and one needs to work with the full Hamiltonian  and Eq.\ (\ref{supermatrix}). 
This break down is seen numerically as 
a fragmentation and disappearance of the $\omega\sim2\Delta$ pair vibration mode for $G/G_c\sim1.2$.  
Eventually, when $\xi_{BCS}\ll R$ or equivalently $G\gg G_c$, 
the lowest collective excitations are not the pair vibrations corresponding to fluctuations in $|\Delta|$
but the Goldstone modes corresponding to \emph{phase} fluctuations 
of $\Delta$~\cite{BruunMottelson,Baranov}. The lowest monopole 
mode is $\omega=2\omega_T$ in this regime.

Let us discuss the open shell case $\mu_F=(n+3/2)\omega_T$ with $n_F=50$. In fig.\ \ref{PairVibfig} (b),
we plot the lowest monopole excitation energy as a function of the coupling strength $G/\omega_T$. 
The $\times$'s are obtained from a numerical calculation of the 
matrix given in Eq.~(\ref{supermatrix}) and the line is Eq.~(\ref{OpenRPA}). 
We see that there is good agreement between the
numerical results and Eq.~(\ref{OpenRPA}) for $G\rightarrow0$. This shows that in the open shell case,
the lowest excitations are essentially quasi-particle excitations with energy $2\Delta$ in the 
intrashell regime. The exact mode frequency is somewhat lower than $2\Delta$ as expected reflecting the residual 
interaction between the quasiparticles. 
With increasing coupling, the intrashell eventually ansatz breaks down for $G\ln(N_F)\sim 0.4$~\cite{BruunHeisel}
making Eq.~(\ref{EffectiveH}) and thus Eq.~(\ref{OpenRPA}) invalid. Again, for stronger coupling when $\xi_{BCS}\ll R$
the Goldstone modes are the lowest excitations.

Several other methods to observe the superfluid transition for atomic Fermi gases have been suggested in the 
literature. They include laser probing of the pairing energy, looking at the expansion properties of the gas, 
and probing the density profile~\cite{Othermethods}.

We have up till now ignored the effects of the Hartree field. This
field introduces a dispersion on the Harmonic oscillator levels with respect to the angular momentum $l$ lifting
the shell degeneracy $\Omega_n=(n+1)(n+2)/2$.
One should therefore replace $\xi_n$ with an $l$-dependent  $\tilde{\xi}_{nl}$ 
in Eq.~(\ref{EffectiveH}), were  $\tilde{\xi}_{nl}$ includes the effect of the Hartree 
field on the single particle energies. The resulting effective Hamiltonian  
describes the pairing correlations in the intrashell regime including the effects of the Hartree field. 
When the splitting introduced by the Hartree field is small compared with the pairing energy $\Delta$, it can be 
ignored. This  is the case for systems with $n_F\lesssim 30$ (depending on the coupling strength)~\cite{BruunHeisel}.
 For very large systems with $n_F\gg200$, there is only 
pairing within the levels with $\tilde{\xi}_{nl}=0$ and $m=-l,-l+1\ldots,l$.
 An analysis completely similar to the one given above yields the lowest excitation  
energy $\omega\simeq2\Delta_{n_Fl}$ in this limit with $\Delta_{n_Fl}$ denoting the pairing energy in the 
level $\tilde{\xi}_{nl}=0$. We plot 
in fig.\ \ref{PairVibfig} (b) as $+$'s the results for lowest monopole excitation energy obtained from a numerical 
calculation based on Eq.~(\ref{supermatrix}) including the effects of the 
Hartree field. For  comparison with the results obtained neglecting the Hartree field, the number of 
particles is fixed in both cases. We see that for $G\rightarrow0$, the Hartree
field suppresses the pairing energy as it introduces a splitting of the
levels in the shell at the chemical potential such that the density of
states for $\tilde{\xi}_{nl}=0$ is reduced. For stronger coupling however, the Hartree field increases the 
pairing energy. This is because the Hartree field increases the density and thus the spatial overlap between 
particles  in the center of trap~\cite{BruunHeisel}. From fig.\ \ref{PairVibfig} (b), 
we see that Eq.\ (\ref{OpenRPA}) reproduces qualitatively the energy of the lowest mode as a function of $G$ even when the 
Hartree field is included. 
 This is as expected since $n_F=50$ is an intermediate value where the
pairing is not in the limit where the Hartree field can be ignored, but also far away from the regime, where the Hartree 
splitting is much larger than the pairing energy.

It should be noted that the  conclusion that there are pair vibration modes 
with an energy $\sim2\Delta$ is independent of the symmetry of the trapping potential. This result
simply reflects the energy required to break individual Cooper pairs. However, the exact value 
of $\Delta$ as a function of the coupling strength depends on the symmetry of the potential and 
Eq.\ (\ref{ARmodesGc}-\ref{ARmodesG0}),(\ref{HogaasenGc}-\ref{OpenRPA}) are of course only valid for a spherical trap.

In summary, we examined in detail the low energy monopole modes of a trapped fermi gas as a function 
of the coupling strength and the chemical potential. Using an effective Hamiltonian describing the pair 
correlations in the intrashell regime we derived 
analytical results that agree well with numerical calculations. As the gas approaches the superfluid 
instability with increasing coupling strength for the closed shell case, the lowest excitation energy corresponding 
to the creation of a Cooper molecule was shown to vanish. 
In the superfluid phase, the lowest  excitation energy is $\omega\simeq2\Delta$
independent of the position of the 
chemical potential. The results presented should be relevant for the present experimental effort to observe the 
superfluid transition. 

We acknowledge valuable discussions with B.\ R.\ Mottelson.

$^*$ Present address: Niels Bohr Institute,  Blegdamsvej 17, 2100 Copenhagen, Denmark

\end{document}